\documentstyle[floats,twocolumn,aps,psfig,prb]{revtex}

\begin{document}
\draft

%***********    This is for two columns *******************************
\twocolumn[\hsize\textwidth\columnwidth\hsize\csname @twocolumnfalse\endcsname

\title{Coherent vs incoherent interlayer transport in layered metals}

\author{
J. Wosnitza$^1$\cite{newadr}, J. Hagel$^1$,
J. S. Qualls$^{2,}$\cite{address}, J. S. Brooks$^2$,
E. Balthes$^3$, D. Schweitzer$^3$,
J. A. Schlueter$^4$, U.~Geiser$^4$,
J. Mohtasham$^5$, R.~W.~Winter$^5$, and G. L. Gard$^5$
}

\address{
 $^1$Physikalisches Institut, Universit\"at Karlsruhe,
  D-76128 Karlsruhe, Germany\\
 $^2$National High Magnetic Field Laboratory, Florida State University,
  Tallahassee, Florida 32306\\
 $^3$3. Physikalisches Institut, Universit\"at Stuttgart,
  D-70550 Stuttgart, Germany\\
 $^4$Materials Science Division,
  Argonne National Laboratory, Argonne, Illinois 60439\\
 $^5$Department of Chemistry, Portland State University,
  Portland, Oregon 97207
}

\date{\today}
\maketitle

\begin{abstract}
The magnetic-field, temperature, and angular dependence of the interlayer
magnetoresistance of two different quasi-two-dimensional (2D) organic
superconductors is reported. For $\kappa$-(BEDT-TTF)$_2$I$_3$ we find
a well-resolved peak in the angle-dependent magnetoresistance at
$\Theta = 90^\circ$ (field parallel to the layers). This clear-cut
proof for the coherent nature of the interlayer transport is absent
for $\beta$''-(BEDT-TTF)$_2$SF$_5$CH$_2$CF$_2$SO$_3$. This and the
non-metallic behavior of the magnetoresistance suggest an incoherent
quasiparticle motion for the latter 2D metal.
\end{abstract}
\pacs{PACS numbers: 74.70.Kn, 72.15.Gd}
% body of paper here
% ***********    This is for two columns *******************************
\vskip2pc]

The usual fundamental concept describing the electronic transport
in metallic crystals is based on the coherent motion of electrons
in band or Bloch states. For a number of cases, however, the
simple semiclassical Boltzmann transport theory fails and a
more complex transport mechanism has to be invoked. \cite{mck98}
Renowned examples are, besides the cuprate superconductors, some
quasi-one-dimensional (1D) \cite{str94,dan95,bali99} and
quasi-two-dimensional (2D) organic metals \cite{wos01} revealing
non-Fermi-liquid properties. Certain signatures in their interlayer
transport suggest an incoherent motion of the charge
carriers between the layers. Incoherent interlayer transport
is expected when the {\it intra}layer scattering rate $\tau^{-1}$ is much
larger than the interlayer hopping integral $t_c$ ($\hbar/\tau \gg
t_c$). In that case the interlayer conductivity is proportional
to the tunneling rate between two adjacent layers and a Fermi
surface is only defined within the layers. \cite{mck98} Nevertheless,
in case the intralayer momentum is conserved during the tunneling
process certain metallic properties persist even without a 3D Fermi
surface.

Some potential candidates which might fit into the above scenario are 
the 2D organic metals and superconductors of the type (BEDT-TTF)$_2X$,
where BEDT-TTF is bis\-ethylene\-dithio-tetrathiafulvalene and $X$
stands for a monovalent anion. Although the observation of magnetic
quantum oscillations provides definitive evidence for a well-developed
2D Fermi surface, \cite{mck98,wosbook} the interlayer transport in
some of these 2D conductors might be incoherent. There exists no
unequivocal proof for an {\it incoherent} transport mechanism as proposed
in Ref.\ \onlinecite{mck98}. There are, however, unambiguous tests for
{\it coherent} interlayer transport: (i) beats in magnetic quantum
oscillations, (ii) a peak in the angular-dependent magnetoresistance
when the magnetic field is parallel to the layers, and (iii) a
crossover from a linear to quadratic field dependence of the
interlayer magnetoresistance. \cite{mck98} These features,
therefore, can experimentally be utilized to preclude incoherent
interlayer transport. Further on, the quantitative analysis of
the features (i) and (ii) can be used to ``measure'' the
degree of two dimensionality, i.e., the value of $t_c$, in
layered metals.

Indeed, in a number of 2D organic conductors the occurrence of
feature (i) and/or (ii) proved the coherent nature of interlayer
transport. \cite{sin01} For the organic metals investigated here,
$\kappa$-(BEDT-TTF)$_2$I$_3$ and
$\beta$''-(BEDT-TTF)$_2$\-SF$_5$CH$_2$\-CF$_2$SO$_3$, feature (i)
is absent, i.e., no beats were detected in magnetic quantum oscillations
down to very low fields, \cite{bal96,bal99,wos00} which render
them possible candidates for metals with incoherently coupled layers.
Although some further aspects of their transport properties could not
be explained by the usual Fermi-liquid theory, \cite{wos01,bal96}
results of the other tests (ii) and (iii) have not been reported so
far. Here we show that a well-defined 3D Fermi surface exists in
$\kappa$-(BEDT-TTF)$_2$I$_3$ whereas no indication for a coherent
interlayer transport can be detected in
$\beta$''-(BEDT-TTF)$_2$\-SF$_5$CH$_2$\-CF$_2$SO$_3$. Indeed, for
the latter material the experimental results clearly reflect
properties which are not explicable by conventional Fermi-liquid
theory.

Since both metals investigated here are superconductors
with $T_c = 3.5$\,K [$\kappa$-(BEDT-TTF)$_2$I$_3$] and $T_c =
4.4$\,K [$\beta$''-(BEDT-TTF)$_2$\-SF$_5$CH$_2$\-CF$_2$SO$_3$],
\cite{bulktc} sufficiently large magnetic fields have to be
applied to attain the normal state for all field orientations.
The band-structure parameters of both metals have been measured
comprehensively by use of de Haas--van Alphen (dHvA) and
Shubnikov--de Haas (SdH) measurements. \cite{bal96,bal99,wos00}
The waveform and the field dependence of the magnetic quantum
oscillations could not be described by 3D theories which proved
both materials as highly 2D metals. \cite{har98,wos00b}
The in-plane Fermi surfaces have been mapped out in detail
utilizing angular-dependent magnetoresistance oscillations (AMRO).
\cite{hel95,wos99} The origin of these oscillations were first
explained by Yamaji \cite{yam89} assuming a corrugated 3D Fermi-surface
cylinder. If this corrugation ($\propto t_c$) indeed exists and if
it is large enough, beats of the magnetic quantum oscillations are
expected. The absence of these beats sets an upper limit for $t_c$
(see below). However, since a 3D Fermi surface is not a necessary
ingredient to explain AMRO, \cite{mck98} the specification of a
corrugation by $t_c$ might be meaningless; incoherent interlayer
transport might be present instead.

The single crystals investigated in this study have been prepared
electrochemically as described earlier. \cite{ben84,gei96}
Thin current leads (15\,$\mu$m gold wire) were glued with
graphite paste to the samples. The interplane resistance was
measured by a four-point method with a current of a few $\mu$A
either by use of a low-frequency ac-resistance bridge or a
lock-in amplifier. The measurements were performed at the High
Magnetic Field Laboratory in Tallahassee in a dilution refrigerator
equipped with a superconducting 20\,T magnet and in a $^3$He
cryostat in fields up to 33\,T. Thereby, the samples could be
rotated {\it in situ} around one axis.

\begin{figure}[ht]
  \centerline{\psfig{file=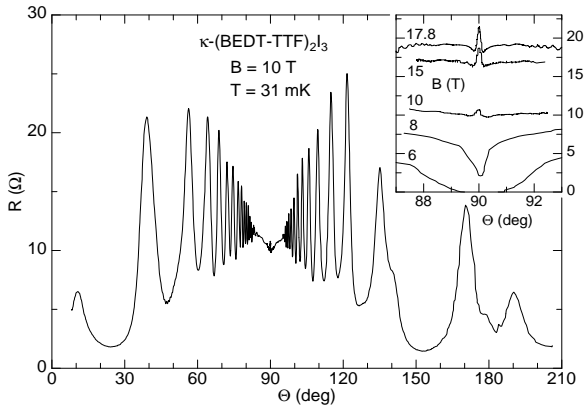,clip=,width=8cm}}
\caption[]{Angular dependence of the interlayer resistance of
$\kappa$-(BEDT-TTF)$_2$I$_3$ at $B = 10$\,T and $T = 31$\,mK.
The inset shows the region close to 90\,deg for different
magnetic fields. At $B \ge 10$\,T a clear resistance peak
evolves at 90\,deg.}
\label{rvsang}
\end{figure}

Figure \ref{rvsang} shows the angular dependence of the resistance
of $\kappa$-(BEDT-TTF)$_2$I$_3$ measured at $B = 10$\,T and
$T = 31$\,mK. The huge oscillations - $R$ changes by more than
a factor of 10 - are found to be equidistant in $\tan\Theta$,
where $\Theta$ is the angle between the applied magnetic field
$B$ and the normal to the conducting plane. Similar AMRO data
with smaller amplitude at $T = 1.6$\,K have been reported earlier.
\cite{hel95} The maxima of the oscillations are given by
\begin{equation}
\tan\Theta = \frac{\pi (n \pm 1/4)}{k_B^{max} c'},
\end{equation}
where $n$ counts the maxima, $c'$ is the spacing between adjacent
layers, and $k_B^{max}$ is the maximum projection of the in-plane
Fermi wave vector $k_F(\varphi)$ onto the field-rotation plane.
$\varphi$ is an azimuthal angle. The minus (plus) sign corresponds
to positive (negative) angles. This simplified formula is valid when
no in-plane component of the hopping vector exists. \cite{kar92}
Here, the linear regression of the peak number $n$ vs $\tan\Theta$
yields $k_B^{max} = 3.36(5)\times 10^{9}$\,m$^{-1}$ with $c' =
1.64$\,nm for $\kappa$-(BEDT-TTF)$_2$I$_3$. This agrees with the
result of Ref.\ \onlinecite{hel95} and fits the assumption of an
almost circular in-plane Fermi surface with $k_F =
k_B^{max} =$ const. In that case $k_F$ is given by $k_F =
(2eF/\hbar)^{1/2} = 3.43\times 10^9$\,m$^{-1}$, with the well-known
dHvA frequency of the so-called $\beta$ orbit of $F = 3870$\,T.
\cite{wosbook,bal96,bal99,har98,hel95}

As mentioned, the bare observation of an AMRO signal is no proof
for a 3D Fermi surface. Indeed, for $\kappa$-(BEDT-TTF)$_2$I$_3$
no nodes in the dHvA and SdH signals are visible with oscillations
of the $\beta$ orbit starting at about $B_{min} = 2.8$\,T. \cite{bal99}
This means that the maximum dHvA-frequency difference is $\Delta F =
(3/4)B_{min} = 2.1$\,T. \cite{bmin} Consequently, the estimated
corrugation amplitude should be less than $t_c \approx 16\,\mu$eV,
since $\Delta F/F = 4t_c/\epsilon_F$ with the Fermi energy $\epsilon_F
= \hbar^2k_F^2/ 2m^*$ and the effective mass $m^*$ ($= 3.9\,m_e$ for
$\kappa$-(BEDT-TTF)$_2$I$_3$.) \cite{wosbook,bal96}) This maximum
$t_c$ is indeed much smaller than $\hbar/\tau \approx 0.14$\,meV
estimated from a Dingle temperature of about 0.25\,K corresponding
to a scattering time $\tau \approx 4.9\times 10^{-12}$\,s.
\cite{bal96,hel95} Therefore, according to the so-called
Mott-Ioffe-Regel incoherent interlayer transport
might be expected. However, looking carefully at the resistance
data around 90$^\circ$ (Fig.\ \ref{rvsang}) a small peak can be seen
in $R$. This becomes much clearer in the inset of Fig.\ \ref{rvsang}
where data taken with high angular resolution are shown for different
magnetic fields at angles close to $90^\circ$. As soon as
the superconductivity is quenched completely, the peak at 90$^\circ$
evolves and becomes larger in amplitude at higher fields. This peak
definitely proves that the interlayer transport in
$\kappa$-(BEDT-TTF)$_2$I$_3$ is {\it coherent}.
This and previous results \cite{sin01} indicate that the $\tau$
obtained from a Dingle analysis seems to have no relation
to the relevant scattering time in the Mott-Ioffe-Regel. Therefore,
this {\it Regel} should only be used as an order-of-magnitude
estimate for possible incoherent transport.

As observed previously for other 2D materials small local minima
to the left and right of the $90^\circ$ peak evolve.
\cite{kar92,ohm99,han98} The peak itself is very narrow with a
full width between the minima of only about $0.34^\circ$,
independent of field strength. Although this is much narrower than
reported for any other 2D metal so far, it is broader than expected
from the maximum $t_c$ estimated above. Although there is a dispute
on whether the physical origin of the $90^\circ$ peak is due to
self-crossing orbits \cite{pes99} or due to small closed orbits,
\cite{han98} there is no controversy that the peak occurs only for
a 3D warped Fermi surface. Assuming a symmetric cylindrical
Fermi-surface topology, Hanasaki {\it et al.} have derived a
relation between the Fermi-surface parameters and the half width of
the peak. \cite{han98} By use of their equation $\Theta_{peak/2} =
t_c c'k_F/\epsilon_F$ with $\Theta_{peak/2} = 0.17^\circ$ in our case,
we obtain $t_c \approx 61\,\mu$eV which is about a factor of four
larger than the maximum $t_c$ estimated from the absence of beating
nodes. This difference cannot be explained by an azimuthal, i.e.,
$\varphi$ dependence of $t_c$ as observed for Sr$_2$RuO$_4$.
\cite{ohm99,ohm00,ber00} Careful AMRO measurements of another
$\kappa$-(BEDT-TTF)$_2$I$_3$ sample for different $\varphi$
resulted -- consistently within error bars -- in $\Theta_{peak/2}
= 0.20(2)^\circ$ independent of azimuthal angle.
Our results indicate that the theories need to be refined for a
quantitatively better estimate of $t_c$.

\begin{figure}[ht]
  \centerline{\psfig{file=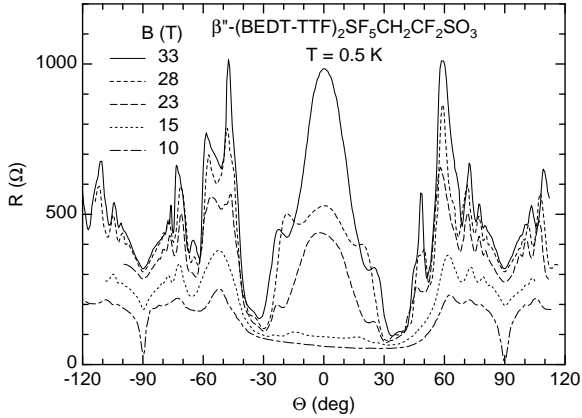,clip=,width=8cm}}
\caption[]{Angular dependence of the interlayer resistance of
$\beta^{''}$-(BEDT-TTF)$_2$SF$_5$CH$_2$CF$_2$SO$_3$ at $T = 0.5$\,K
for different magnetic fields up to 33\,T.}
\label{sf5rvsan}
\end{figure}

A {\it qualitatively} different picture occurs for the 2D organic
metal $\beta^{''}$-(BEDT-TTF)$_2$SF$_5$CH$_2$CF$_2$SO$_3$.
From the absence of beating nodes in dHvA and SdH oscillations
that start at about $B_{min} = 1.7$\,T with a frequency
of $F = 199(1)$\,T and a cyclotron effective mass of $m^* =
2.0(1)\,m_e$, \cite{wos00} we estimate a maximum $t_c =
18.5\,\mu$eV. Previous AMRO measurements showed that the small
Fermi surface -- occupying only 5\% of the in-plane Brillouin zone --
consists of a strongly elongated ellipsoid with an axis ratio of
about 1:9 ($0.26\times 10^9$\,m$^{-1} < k_F < 2.4\times 10^9$\,m$^{-1}$).
\cite{wos99} In this experiment the applied field of 10\,T was not
sufficient to suppress superconductivity at $90^\circ$. At $T =
0.5$\,K, fields above about 15\,T are necessary to reach the normal
state. However, as Fig.\ \ref{sf5rvsan} shows, even for fields up
to 33\,T no indication of a peak at $90^\circ$ appears. With increasing
field only AMRO peaks and SdH oscillations become dominant. \cite{peak0}
From a linear regression of the AMRO peak number vs $\tan\Theta$ we
obtain $k_B^{max} \approx 1.1\times 10^9$\,m$^{-1}$ with $c' =
1.74$\,nm.

\begin{figure}[ht]
  \centerline{\psfig{file=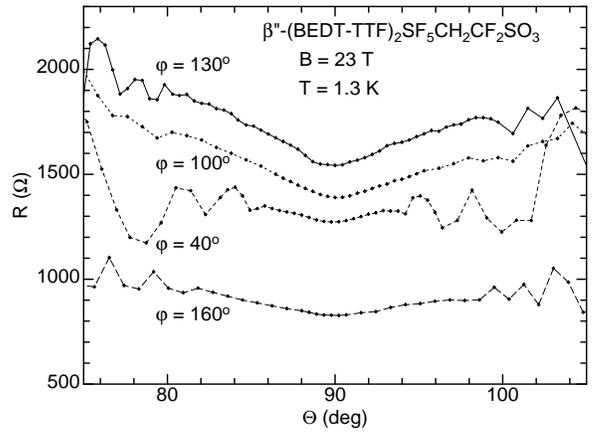,clip=,width=8cm}}
\caption[]{Interlayer resistance close to 90\,deg of another
$\beta^{''}$-(BEDT-TTF)$_2$SF$_5$CH$_2$CF$_2$SO$_3$ sample
for different azimuthal angles $\varphi$.}
\label{sf5nopk}
\end{figure}

The data shown in Fig.\ \ref{sf5rvsan} were taken at fixed
azimuthal angle $\varphi \approx 80^\circ$, where $\varphi = 0$
corresponds to a field rotation through the $k_a$ axis.
\cite{wos99} Since $t_c$ may vary largely with $\varphi$,
additional AMRO data were collected at a number of different
$\varphi$. Figure \ref{sf5nopk} shows the resistance of a second
sample for four different $\varphi$ at $T = 1.3$\,K for $B = 23$\,T
close to $\Theta = 90^\circ$. For all investigated azimuthal angles
$\varphi$, all magnetic fields, and all samples never a peak at
$90^\circ$ occured. With an approximate angular resolution of
$0.01^\circ$ for the polar angle $\Theta$ ($R$ was continuously
monitored when the samples were rotated manually), $t_c$ must
be smaller than about $10^{-6}$\,eV estimated conservatively
by use of above formula for $\Theta_{peak/2}$. This almost two
orders of magnitude smaller $t_c$ than observed so far
strongly suggests an {\it incoherent} interlayer-transport
mechanism for $\beta^{''}$-(BEDT-TTF)$_2$SF$_5$CH$_2$CF$_2$SO$_3$.

\begin{figure}[ht]
  \centerline{\psfig{file=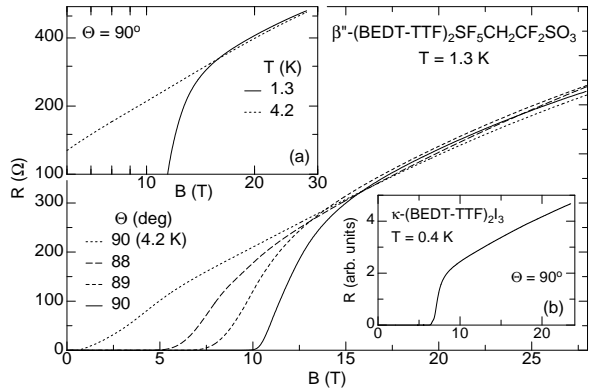,clip=,width=8cm}}
\caption[]{Field dependence of the interlayer resistance of a third
$\beta^{''}$-(BEDT-TTF)$_2$SF$_5$CH$_2$CF$_2$SO$_3$ sample close
to 90\,deg. The inset (a) shows the data for $\Theta = 90$\,deg at
$T = 1.3$ K and $T = 4.2$\,K in a double-logarithmic scale. The
inset (b) shows $R$ vs $B$ for $\kappa$-(BEDT-TTF)$_2$I$_3$ at
90\,deg.}
\label{rvsb90}
\end{figure}

As a final test for coherent transport [see point (iii) above],
we measured carefully the field dependence of the interlayer
resistance for fields aligned within the highly conducting planes.
Perfect alignment of the samples was easily achieved in fields
low enough to retain superconducting traces at $\Theta = 90^\circ$.
The resulting data (Fig.\ \ref{rvsb90}) show clearly that $R$ at
$\Theta = 90^\circ$ grows less than linear with $B$. For intentionally
misaligned field orientations a somewhat steeper, but still less
than linear field dependence is observed (see the examples at
$88^\circ$ and $89^\circ$ in Fig.\ \ref{rvsb90}). Thus, there
is definitely {\it no} indication for a crossover to quadratic
behavior in $B$ as expected for coherent transport at large fields.
\cite{mck98,schof00} There is, however, a drawback regarding the
relevance of this test: for $\kappa$-(BEDT-TTF)$_2$I$_3$, we find
almost the same field dependence of $R$ at $90^\circ$ [inset (b) of
Fig.\ \ref{rvsb90}] although the peak at $90^\circ$ proves coherent
transport. Equally, for the layered metal Sr$_2$RuO$_4$ not a $B^2$
behavior but a superlinear $B$ dependence ($\propto B^{1.5}$)
was observed. \cite{ohm00} Although $eB\tau/m^* \gg 1$ seems to be
fulfilled, larger fields might be necessary to verify the $B^2$
behavior. \cite{schof00} A double-logarithmic plot of $R$ vs $B$
[inset (a) of Fig.\ \ref{rvsb90}] reveals that in the present case
$R$ grows approximately with $B^{0.9}$ at $T = 4.2$\,K. However,
both in the linear as well as in the double-logarithmic plot
clear curvatures of the data are apparent. Above about 20\,T,
$R \propto \ln B$ fits the data reasonably well (not shown). However,
higher fields are necessary to determine the limiting field dependence.

All the above-discussed results give {\it no} experimental evidence
for coherent interlayer transport in the 2D organic metal
$\beta^{''}$-(BEDT-TTF)$_2$SF$_5$CH$_2$CF$_2$SO$_3$. Along these lines,
previous results corroborate the existence of only very weakly coupled
perfectly two-dimensional metallic sheets with non-Boltzmann-like
interlayer transport. Accordingly, the pronounced two dimensionality
of the Fermi surface is evidenced by inverse-sawtooth-like dHvA
oscillations which perfectly fit the theoretical prediction for a 2D
metal with fixed chemical potential. \cite{wos00b} Further on, deviations
from the conventional Bloch-Boltzmann transport theory were observed in
the interlayer magnetoresistance for fields close to $\Theta = 0$.
\cite{wos01} A field-induced metal-insulator transition and a violation
of Kohler's rule was found. \cite{nam01} All these peculiarities reflect
that $\beta^{''}$-(BEDT-TTF)$_2$SF$_5$CH$_2$CF$_2$SO$_3$ is a highly
unusual metal. On the one hand, the interlayer resistance at $B = 0$
is metallic from lowest $T$ up to room temperature for all samples
we investigated and a 2D in-plane Fermi surface can clearly be resolved.
On the other hand, the electronic transport perpendicular
to the layers is most probably incoherent and cannot be described
by conventional theories.

In conclusion, we proved that the highly 2D organic metal
$\kappa$-(BEDT-TTF)$_2$I$_3$ has a well-developed 3D Fermi
surface and the electronic transport can be described by the
coherent motion of electrons in Bloch states.
The interlayer overlap integral $t_c \approx 61\,\mu$eV is only
slightly smaller than the scattering rate $\hbar/\tau
\approx 0.14$\,meV setting this material just at the borderline to
incoherent electronic transport. The latter seems to occur in the
organic metal $\beta^{''}$-(BEDT-TTF)$_2$SF$_5$CH$_2$CF$_2$SO$_3$
for which $t_c < 1\,\mu$eV and
where all experimental tests to observe signatures for coherent
interlayer transport failed.

We thank R.\ H. McKenzie for helpful comments and N.\ Tiedau for
performing some supplementary transport measurements.
Work at Karlsruhe was supported by the DFG.
J.S.Q.\ was supported by Grant No.\ NSF-DMR-10427. We acknowledge
the NSF, the state of Florida, and
the U.S.\ Dept.\ of Energy for support of the National High Magnetic
Field Laboratory. Work at Argonne National
Laboratory was supported by the U.S.~Dept.\ of Energy (W-31-109-ENG-38).
Work at Portland State University was supported by NSF
(Che-9904316) and the Petroleum Research Fund (ACS-PRF No.\ 34624-AC7).


\end{document}